\documentstyle[11pt,newpasp,twoside,epsf]{article}
\def\edcomment#1{\iffalse\marginpar{\raggedright\sl#1\/}\else\relax\fi}
\reversemarginpar

\newcommand{\HI}{\protect\normalsize H\thinspace\protect\footnotesize
I\protect\normalsize} 
 
\newcommand{\HIb}{\protect\normalsize\bf H\thinspace\protect\footnotesize\bf
I\protect\normalsize\bf}

\newcommand{\ij}{\mbox{$I\!-\!J$}}
\newcommand{\jk}{\mbox{$J\!-\!K$}}
\newcommand{\B}{{$B$}}

\newcommand{\II}{{$I$}}
\newcommand{\J}{{$J$}}
\newcommand{\HH}{{$H$}}
\newcommand{\K}{{$K_s$}}
\newcommand{\tfr}{Tully\,--\,Fisher relation}

\newcommand{\kms}{\,km\,s$^{-1}$}
\newcommand{\etal}{et~al.\ }
\begin{document}
\title{Multi-wavelength Observations of Galaxies in the Southern Zone of Avoidance }
 \author{Anja Schr\"oder }
\affil{Observatoire de la C\^ote d'Azur, B.P.4229, 06304 Nice Cedex 04, France }
\author{Ren\'ee C. Kraan-Korteweg }
\affil{Depto. de Astronom\'{\i}a, Universidad de Guanajuato,
 Apdo. Postal 144, Guanajuato GTO 36000, Mexico }
\author{Gary A. Mamon }
\affil{IAP, 98~bis Blvd Arago, 75014 Paris, France, and \\
DAEC, Observatoire de Paris-Meudon, 92195 Meudon, France}

\begin{abstract}
We discuss the possibilities of extragalactic large-scale studies behind the
Zone of Avoidance (ZOA) using complementary multi-wavelength data from
optical, systematic blind \HI , and near-infrared (NIR) surveys. Applying these
data to the NIR \tfr\ permits the mapping of the peculiar velocity field
across the ZOA. Here, we present results of a comparison of galaxies
identified in the rich low-latitude cluster Abell\,3627 in the \B -band with
NIR (DENIS) data, and cross-identifications of galaxies detected
with the blind Parkes \HI\ Multibeam survey with NIR data -- many of 
which are optically invisible.
\end{abstract}

\section{Introduction}

Understanding the origin of the peculiar velocity of the Local Group and the
dipole in the Cosmic Microwave Background is one of the major goals of the
study of large-scale structures. Reconstructions of large-scale structures,
for instance, still suffer from large interpolation uncertainties across the
Zone of Avoidance (ZOA), which extends over about 25\% of the optically
visible extragalactic sky.  Dynamically important structures might still lie
hidden in this zone, such as the recently discovered nearby galaxy Dwingeloo~1 
(Kraan-Korteweg \etal 1994) and the rich massive cluster Abell~3627
(Kraan-Korteweg \etal 1996). Important large-scale structures, e.g., the
Supergalactic Plane and other filaments and wall-like structures, seem to
continue across this zone. A more complete knowledge of the distribution of
galaxies in redshift space, as well as in distance space, will improve the
reconstructed galaxy density fields and help to explain the origin of the
peculiar velocity of the Local Group and the dipole in the Cosmic Microwave
Background.

Various approaches are presently being employed to uncover the galaxy
distribution in the ZOA (cf.\ Kraan-Korteweg \& Lahav 2000): deep optical
searches, NIR surveys (DENIS and 2MASS), far-infrared surveys (e.g., IRAS),
and blind \HI\ searches.  All methods produce new results, but all suffer from
limitations and selection effects.  The combination of data from an optical
galaxy search, a NIR survey and a systematic blind \HI\ survey will allow us
to examine the large-scale structures behind the southern Milky Way and the
peculiar velocity field associated with them.  Redshift independent distance
estimates can be obtained via the NIR \tfr . Bouch\'e \& Schneider (these
proceedings) have shown that the \K -band is ideal for this approach.

Here we use (i) data from the diameter-limited, deep \B -band galaxy
survey by Kraan-Korteweg and collaborators for cross-identifications at
intermediate extinctions, (ii) spiral galaxies detected with the systematic
blind \HI\ survey of the southern ZOA ($|b| \la 5\deg$) with the Parkes
Multibeam (MB) receiver, plus pointed \HI\ observations of partially obscured
galaxies at intermediate latitudes $5\deg\!< |b| <\!10\deg$ (Kraan-Korteweg
et~al.\ 2000), and (iii) the DENIS survey:

(i) The deep optical survey in the southern ZOA is being conducted by one of
us (cf. Kraan-Korteweg \&~Woudt 1994, Kraan-Korteweg 2000, Woudt et~al., these
proceedings). In this region ($265\deg \la \ell \la 340\deg$, $|b| \la
10\deg$), over 11\,000 previously unknown galaxies above a diameter limit of
$D\!=\!0\farcm2$ have been identified, next to the previously known
$\sim\!300$ Lauberts galaxies with $D \ge 1^{\prime}$ (Lauberts 1982). As
shown by Kraan-Korteweg (2000), this diameter limited survey is complete to $A_B
\simeq 3^{\rm m}$, although galaxies can be identified to \mbox{$A_B \la
5^{\rm m}$} (or $|b| \simeq 5\deg$ on average).

(ii) The Multibeam (MB) ZOA survey is a systematic \HI\ survey of the southern
ZOA within $|b| \la 5\deg$ (see Henning et al., Staveley-Smith et al., these
proceedings). It will trace gas-rich spirals out to redshifts of
12\,000\,\kms\ with no hindrance from the Galactic dust. The survey is being
conducted with the Multibeam receiver (13 beams) at the 64\,m Parkes telescope
and should detect of the order of 1500 galaxies above the $5 \sigma$ detection
limit of 10\,mJy. Only few of these predicted galaxies will have an optical
counterpart, but many might be visible in the NIR. Their identifications
become feasible with the NIR surveys such as DENIS (Epchtein 1997, Epchtein et
al. 1997) and 2MASS (Skrutskie \etal 1997).

(iii) The DENIS survey has currently imaged about 80\% of the southern sky in
the $I\,(0.8\,\mu)$, $J\,(1.25\,\mu)$ and $K_s (2.15\,\mu)$ passbands with a
resolution of $1\arcsec$ in \II\ and $3\arcsec$ in \J\ and \K . In a pilot
study, we have assessed the performance of the DENIS survey at low Galactic
latitudes (Schr\"oder \etal 1997, hereafter Paper I; Kraan-Korteweg \etal
1998, hereafter Paper II; Schr\"oder \etal 1999, hereafter Paper III). After
giving some details about the DENIS survey, we present improved results on the
photometry of galaxies in the cluster Abell\,3627 (Sect.~3), and on
cross-identifications with galaxies detected in the \HI\ MB survey (Sect.~4).

\section{The DENIS Survey} \label{denis}

Observations in the NIR have several advantages over other ZOA surveys, and
they provide important complementary data. Compared to the optical, the NIR is
less affected by the foreground extinction (the extinction in \K\ is about
10\% of the extinction in \B ). The NIR is sensitive to early-type galaxies,
tracers of massive groups and clusters which are neither uncovered in far
infrared surveys nor in the 21\,cm radiation.  The NIR shows little confusion
with Galactic objects such as young stellar objects and cool cirrus
sources. Moreover, the NIR allows a good estimation of the stellar mass
content of galaxies because recent star formation contributes only little to
the flux at this wavelength. It is hence ideally suited for the application of
the \tfr .

Number counts of galaxies decrease in the ZOA due to the increasing foreground
extinction. This effect depends, however, on wavelength.  Interpolating from
Cardelli \etal (1989), the extinctions in the NIR passbands are
$A_{I_c}\!=\!0\fm45$, $A_J\!=\!0\fm21$, and $A_{K_s}\!=\!0\fm09$ for
$A_B\!=\!1\fm0$. Thus the decrease in number counts as a function of
extinction is considerably slower in the NIR than in the optical.

Figure~1 shows the predicted surface number density of galaxies as a function
of Galactic foreground extinction for the DENIS completeness limits of $I_{\rm
lim}\!=\!16\fm5$, $J_{\rm lim}\!=\!14\fm8$, $K_{\rm lim}\!=\!12\fm0$ (Mamon
1998) and for $B_{\rm lim}\!=\!19\fm0$ (Gardner \etal 1996).  The figure
suggests that the NIR becomes notably more efficient at $A_B \ga 2^{\rm m}$,
that the \J -band is the most efficient passband to find galaxies at
intermediate extinctions, and that \K\ becomes superior to \J\ at $A_B \simeq
12^{\rm m}$.  While the diameter-limited samples of the optical searches
become incomplete at $A_B \simeq 3^{\rm m}$, the \J - and \K -bands will
easily detect galaxies up to $A_B = 10^{\rm m}$, or even higher
extinctions. These are very rough predictions and do not take into account any
dependence on morphological type, surface brightness, orientation and
crowding, which may lower the number of actually detectable galaxies (Mamon
1994).

\begin{figure}[tb]
\vspace{-2.8cm}
\plotone{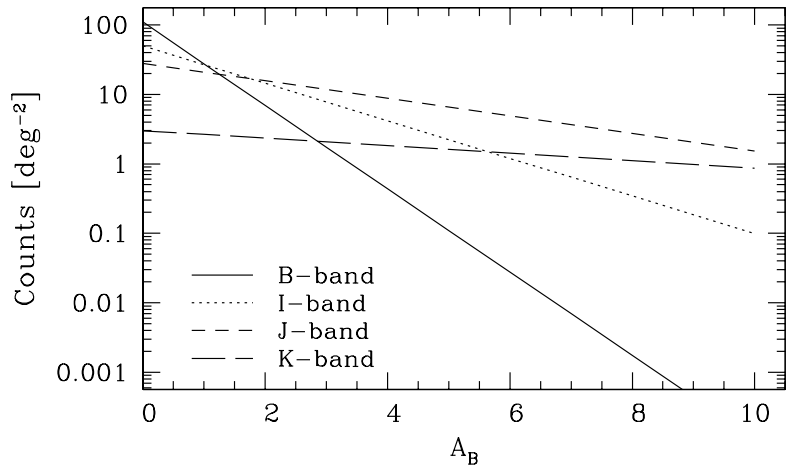}\\
\vspace{-4cm}
\caption{Predicted galaxy counts in \B , \II , \J\ and \K\ as a function of
absorption in \B , for highly complete and reliable DENIS galaxy samples and a
$B_J \leq 19^{\rm m}$ optical sample.  }
\label{galctsplot}           
\end{figure}

\section{NIR Photometry in the Norma Cluster } \label{phot}

We investigated currently available DENIS data at the core of the Great Attractor, i.e., in the
low-latitude ($\ell\!=\!325\deg$, $b\!=\!-7\deg$), rich cluster Abell 3627
(cf.\ Woudt et al., these proceedings), where the Galactic extinction is well
determined (Woudt \etal 1998). Five high-quality DENIS strips cross the
cluster Abell 3627. The inspected 110 images cover about one-fifth of the
cluster area within its Abell-radius of $R_A = 1\fdg75$ (each DENIS image is
$12\arcmin$x$12\arcmin$, offset by $10\arcmin$ in declination and right
ascension). The extinction over the cluster area varies as $0\fm6 \le
A_B \le 2\fm2$.

We cross-identified the galaxies found in the optical survey with the DENIS
\II, \J, and \K\ images.  On the 110 images, 234 galaxies had been identified
in the optical. We have recovered 198 (85\%) galaxies in the \II\ band, 183
(78\%) in the \J\ band, and 123 (53\%) in the \K\ band (not including galaxies
visible on more than one image). At these extinction levels, the optical
survey does remain the most efficient in {\it identifying} obscured
galaxies. In the NIR, the \II - and \J -band are equally efficient, though the
severe star crowding makes identification of faint galaxies difficult in \II .

We have obtained preliminary \II , \J\ and \K\ Kron photometry using the
automated galaxy extraction pipeline (Mamon \etal 1997b) on the galaxies
visually identified by us. Although many of the
galaxies have a considerable number of stars superimposed on their images,
comparison of the magnitudes derived from this fairly automated algorithm
agree well with the few known, independent measurements.

\begin{figure}[tb]
\vspace{-7.8cm}
\plotone{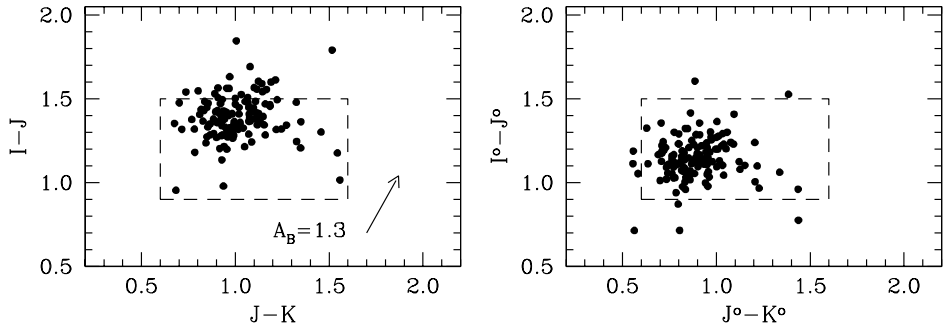} \\
\vspace{-1.cm}
\caption{The colour\,--\,colour diagram for the Norma cluster. The dashed box marks
  the colour range of unobscured galaxies.  Observed colours are displayed in the
  left panel with the arrow indicating the expected mean offset due to
  extinction. The right panel shows extinction-corrected colours. }
\label{photnplot}           
\end{figure}

The NIR magnitudes have been used to study the colour\,--\,colour diagram \ij\
versus \jk\ (Fig.~2). In the left hand panel, observed colours (from a
$7\arcsec$ aperture) are displayed; in the right hand panel the colours are
corrected for foreground extinction using Mg$_2$-indices values and
interpolations according to the Galactic \HI\ distribution. As a comparison,
the range in colours of galaxies at high latitudes (Mamon \etal 1998) is indicated
by the box. The displacement of the points agrees well with the path of
extinction (arrow) based on the mean extinction in these five strips of
$A_B=1\fm3$ (Woudt \etal 1998), suggesting that our preliminary photometry is
reasonably accurate.  Moreover, the shift in colour can be fully explained by
the foreground extinction or, more interestingly, the NIR colours of obscured
galaxies provide, in principle, an independent way of mapping the extinction
in the ZOA (see also Mamon \etal 1997a).

\section{Cross-identification on DENIS
  Images of \HIb  -detected Galaxies  } \label{hi}

\begin{figure}[tb]
\vspace{-6.8cm}
\plotone{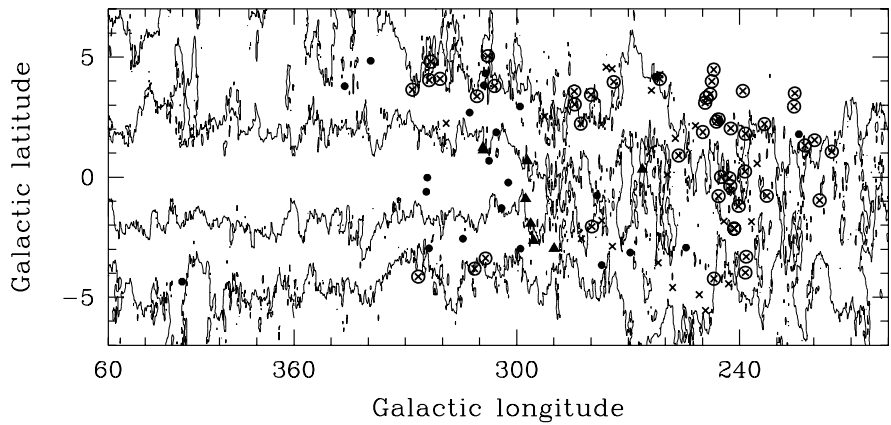}\\
\caption{Distribution of 100 galaxies detected with the shallow MB survey in 
  the southern ZOA. The superimposed contours represent absorption levels
  of $A_B = 3\fm0$ (outer contour) and $10\fm0$ (see text for details). MB
  galaxies detected with DENIS data are marked with a cross (visible in \II )
  or a triangle (not visible in \II ); those with an optical counterpart are displayed
  with a large circle; those not detected in either \B\ or the NIR are marked with
  a filled circle. }
\label{ssdmapplot}           
\end{figure}

Figure~3 displays the distribution of galaxies detected in the shallow MB-ZOA
survey (Henning \etal 2000).  Contours indicate extinction levels determined
from the DIRBE maps (Schlegel \etal 1998).  The outer contour corresponds to
$A_B = 3\fm0$, the completeness limit for galaxies with an
extinction-corrected diameter of $D^{o}\!=1\farcm3$ in deep optical ZOA galaxy
catalogues (see Kraan-Korteweg, these proceedings). The inner contour
indicates $A_B=10^{\rm m}$ (the Milky Way becomes opaque in the optical at $A_B \la
5^{\rm m}$).

For 100 of the 110 galaxies detected in the shallow \HI\ survey, DENIS images
($12\arcmin \times 12\arcmin$) covering the full positional uncertainty region
($\sim\!4\arcmin \times\!\!\sim\!4\arcmin$) were currently available. 77
galaxies can be detected in the NIR: 69, 67 and 58 in \II\ (crosses), \J , and
\K\ respectively, while only 48 have been detected in the \B -band (large
circles in Fig.~3). Triangles indicate the 8 galaxies visible in \J\ and/or
\K\ but not in the \II -band. They are clearly at higher extinction levels
than the galaxies seen in \II\ or \B . This is clear also in the histogram in
Fig.~4 which shows the wavelength-dependence of detection rate (shaded versus
unshaded regions) as a function of foreground extinction. While the detection
rate in the \B -band decreases rapidly with increasing extinction, the
decrease in \II\ and particularly in \K\ is much slower. The \II -band
(together with the \J -band) seems to be the best passband to find galaxies at
extinction levels between $2^{\rm m}$ and $\sim10^{\rm m}$ (cf.\ previous
section), and the \K -band becomes superior at $A_B > 10^{\rm m}$. Keeping in
mind (a) the low number statistics, (b) the fact that MB galaxies are gas-rich 
spiral and irregular galaxies, and (c) that the optical searches are
diameter limited rather than magnitude limited, Fig.~4 and Fig.~1 compare very
well.

\begin{figure}[tb]
\vspace{-0.5cm}
\plotone{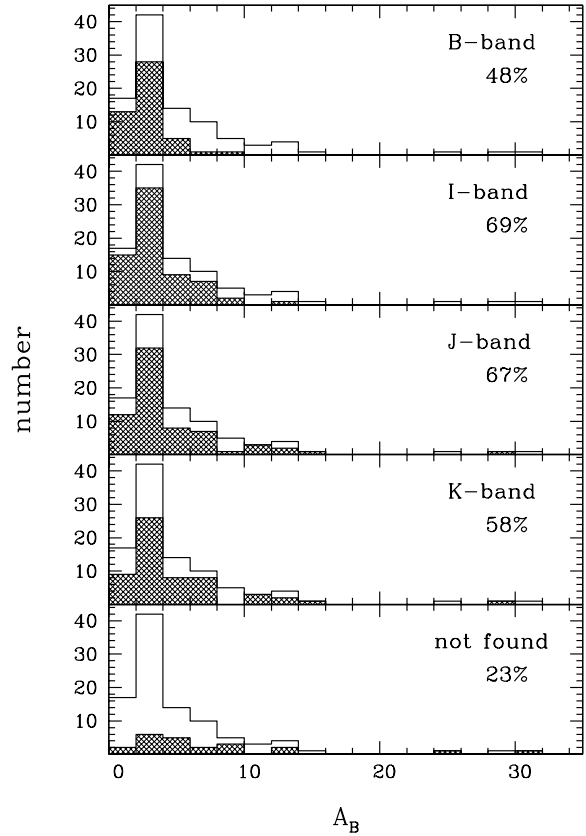}\\
\vspace{-1.5cm}
\caption{Galaxy detections as a function of foreground extinction (shaded
  area). The unshaded region refers to all galaxies detected in the shallow survey. }
\label{sshistplot}           
\end{figure}

For 23 galaxies no counterpart could be found. These galaxies either lie
behind a very thick extinction layer (e.g., one galaxy at $b \simeq 0\deg$ has
$A_B \sim 70^{\rm m}$ according to the DIRBE maps), or they are late-type
galaxies of very low surface brightness, hence below the magnitude limits of
the DENIS survey. The \HI\ survey (unshaded region in Fig.~4), is, however,
not affected by the foreground extinction, therefore superior to other
passbands in uncovering spiral galaxies at low Galactic latitudes and high
foreground extinction levels. The low number rate at high extinctions is
partly due to confusion with Galactic continuum sources at lowest latitudes,
as well as the Local Void (e.g., Henning et al., these proceedings).

Figure~5 shows the dependence of the observed \jk\ colour of these galaxies
(from a $7\arcsec$ aperture) on foreground extinction $A_B$, including data
from the low-latitude cluster Abell 3627 (stars; see previous section).  The
broader scatter for the shallow survey galaxies (and some of their companions)
can be explained by the larger error in the photometry due to the increase in
star crowding. However, the systematic offset towards the red with increasing
extinction (the reddening path is indicated by the arrow) is clearly evident.

\begin{figure}[tb]
\vspace{-4cm}
\plotone{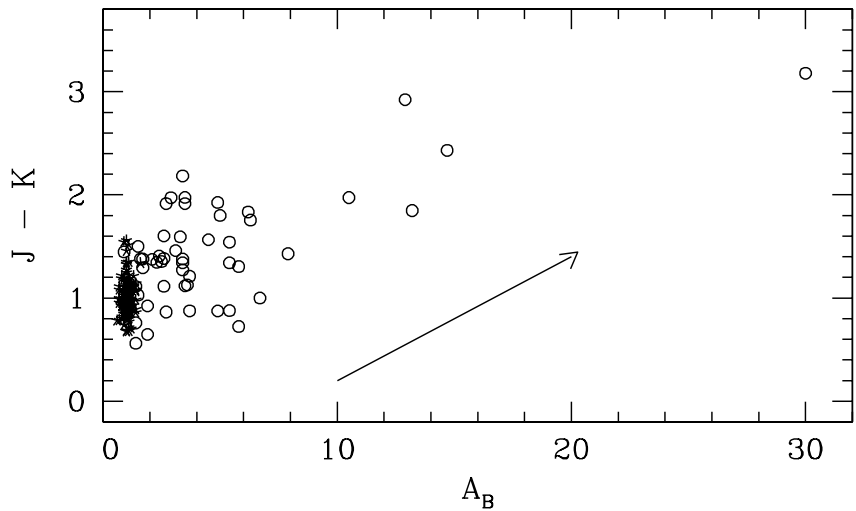}\\
\vspace{-2cm}
\caption{NIR colour versus foreground extinction. Stars refer to
  galaxies in the Norma cluster, while open circles are galaxies found with
  the shallow \HI\ survey. }
\label{ssextplot}           
\end{figure}

\section{Conclusion }

We have demonstrated the potential of a multi-wavelength approach to penetrate
the extinction layers of the Milky Way for studying extragalactic large-scale
structures. The detection rate of the three surveys (a deep optical, a
systematic blind \HI\, and a NIR survey) depends on galaxy type and foreground
extinction.  The three surveys together find galaxies at all extinction levels
and Galactic latitudes. Furthermore, with the available photometry, the NIR
\tfr\ can be applied to most of these galaxies. The latter allows the mapping
of the peculiar velocity field across the whole ZOA.

Optical surveys are superior for identifying galaxies at intermediate
latitudes and extinctions ($|b|\!\!>\!\!5\deg$, $A_B\!\la\!3^{\rm m}$). 
The additional NIR luminosities and colours will prove invaluable in analysing
the optical survey data as well as the distribution of the galaxies in redshift space,
and in the final merging of these data with existing sky surveys.

With the DENIS NIR survey we can trace galaxies down to $A_B \la 15^{\rm
m}$, i.e. very low latitudes. Though the \II -band is strongly affected by
star crowding (which mainly depends on the limiting magnitude of the survey),
it is best suited for identifying galaxies up to $A_B \la 4^{\rm m}$ because
of the higher spatial resolution ($1\arcsec$). At higher foreground
extinctions, the \J -band and the \K -band (for $A_B \ga 10^{\rm m}$) become
important. NIR surveys thus further reduce the width of the ZOA. In
particular, they provide the only tool with which to identify early-type
galaxies at high extinction. Despite the star crowding at these latitudes, \II
, \J\ and \K\ photometry from the survey data can be successfully performed
and the colours can be used to calibrate the DIRBE extinction maps locally.

In addition, we can complement the NIR data-set using the \HH -band and the \K
-band (with a fainter magnitude limit) of 2MASS (see also Huchra et~al., these
proceedings) to obtain a wider range in colours.

The blind \HI\ survey uncovers spiral galaxies independent of
foreground extinction. For a significant fraction of these detections, a DENIS
counterpart has been found. These MB-ZOA data cover the Galactic latitude
range $|b| <\!5\deg$. We will complement this area with pointed \HI\
observations of optically identified spiral galaxies for intermediate
latitudes ($5\deg\!< |b| <\!10\deg$).  About 300 spiral galaxies have already
been detected (Kraan-Korteweg \etal 1997).

\end{document}